\PassOptionsToPackage{unicode}{hyperref}
\PassOptionsToPackage{hyphens}{url}
\documentclass[
]{article}
\usepackage{amsmath,amssymb}
\usepackage{iftex}
\ifPDFTeX
  \usepackage[T1]{fontenc}
  \usepackage[utf8]{inputenc}
  \usepackage{textcomp} 
\else 
  \usepackage{unicode-math} 
  \defaultfontfeatures{Scale=MatchLowercase}
  \defaultfontfeatures[\rmfamily]{Ligatures=TeX,Scale=1}
\fi
\usepackage{lmodern}
\ifPDFTeX\else
\fi
\IfFileExists{upquote.sty}{\usepackage{upquote}}{}
\IfFileExists{microtype.sty}{
  \usepackage[]{microtype}
  \UseMicrotypeSet[protrusion]{basicmath} 
}{}
\makeatletter
\@ifundefined{KOMAClassName}{
  \IfFileExists{parskip.sty}{%
    \usepackage{parskip}
  }{
    \setlength{\parindent}{0pt}
    \setlength{\parskip}{6pt plus 2pt minus 1pt}}
}{
  \KOMAoptions{parskip=half}}
\makeatother
\usepackage{longtable,booktabs,array}
\usepackage{calc} 
\usepackage{etoolbox}
\makeatletter
\patchcmd\longtable{\par}{\if@noskipsec\mbox{}\fi\par}{}{}
\makeatother
\IfFileExists{footnotehyper.sty}{\usepackage{footnotehyper}}{\usepackage{footnote}}
\makesavenoteenv{longtable}
\usepackage{graphicx}
\makeatletter
\newsavebox\pandoc@box
\newcommand*\pandocbounded[1]{
  \sbox\pandoc@box{#1}%
  \Gscale@div\@tempa{\textheight}{\dimexpr\ht\pandoc@box+\dp\pandoc@box\relax}%
  \Gscale@div\@tempb{\linewidth}{\wd\pandoc@box}%
  \ifdim\@tempb\p@<\@tempa\p@\let\@tempa\@tempb\fi
  \ifdim\@tempa\p@<\p@\scalebox{\@tempa}{\usebox\pandoc@box}%
  \else\usebox{\pandoc@box}%
  \fi%
}
\def\fps@figure{htbp}
\makeatother
\providecommand{\tightlist}{%
  \setlength{\itemsep}{0pt}\setlength{\parskip}{0pt}}
\usepackage[table]{xcolor}
\usepackage{booktabs}
\usepackage{array}
\usepackage{longtable}
\usepackage{tabularx}       
\usepackage{etoolbox}

\definecolor{rowgray}{gray}{0.95}
\newcommand{\rowcolorscustom}{\rowcolors{2}{rowgray}{white}}

\makeatletter
\patchcmd{\LT@start}{\@mkpream}{\rowcolorscustom\@mkpream}{}{}
\makeatother

\usepackage{float}
\usepackage{bookmark}
\IfFileExists{xurl.sty}{\usepackage{xurl}}{} 
\hypersetup{
  hidelinks,
  pdfcreator={LaTeX via pandoc}}

\author{}
\date{}

\begin{document}

\section{\texorpdfstring{OLAF: An Open Life Science Analysis Framework
for Conversational Bioinformatics Powered by Large Language
Models}{OLAF: An Open Life Science Analysis Framework for Conversational Bioinformatics Powered by  Large Language Models}}\label{olaf-an-open-life-science-analysis-framework-for-conversational-bioinformatics-powered-by-large-language-models}

\textbf{Authors}

Dylan Riffle {[}1{]}*, Nima Shirooni {[}2{]}, Cody He {[}2{]}, Manush
Murali {[}2{]}, Sovit Nayak {[}2{]}, Rishikumar Gopalan {[}2{]}, Diego
Gonzalez Lopez {[}2{]}

{[}1{]} Weill Cornell Medicine

{[}2{]} University of California, Irvine

{[}*{]} Corresponding author: dyr4005@med.cornell.edu

\section{Summary}\label{summary}

OLAF (Open Life Science Analysis Framework) is an open-source software
platform that leverages large language models (LLMs) to generate and
execute bioinformatics code in response to natural language queries.
Built with a modular agent--pipe--router architecture and a modern web
interface (Angular front-end with a Python/Firebase backend), OLAF
serves as an intelligent lab assistant for computational biology. Users
can ask OLAF to perform complex data analyses or lab automation tasks in
plain English, for example, ``Identify differentially expressed genes in
my single-cell RNA-seq dataset,'' and OLAF's LLM-based agents will write
and run the appropriate code to fulfill the request. This capability
effectively bridges the gap between natural language instructions and
bioinformatics pipelines, empowering researchers who lack advanced
programming skills to leverage state-of-the-art computational tools.

A core innovation of OLAF is its ability to interface with specialized
scientific data formats and tools that general-purpose chatbots cannot
handle. For instance, general LLM systems like ChatGPT operate purely in
the realm of text and have no inherent means to open a complex binary
file (such as a large .h5ad single-cell data file) or execute code on
it. By contrast, OLAF's design allows the LLM agent to directly interact
with data. It can generate Python code using domain libraries (e.g.,
Scanpy {[}1{]} to read an .h5ad file) and then execute it through OLAF's
backend pipeline. The code is run in a sandboxed backend environment,
and the results---whether numerical outputs, plots, or processed
data---are returned to the user through the interactive front-end. This
tightly integrated loop of LLM-driven code generation and automated
execution means OLAF can carry out real bioinformatics analyses
end-to-end, something not achievable with a standalone LLM chat
interface. Importantly, OLAF emphasizes transparency: users can inspect
the generated code and results, helping build trust and enabling
learning by example.

In summary, OLAF provides a rigorous yet accessible platform that
marries cutting-edge LLM capabilities with practical bioinformatics
infrastructure. It lowers the barrier for biologists and clinicians to
perform advanced analyses and lab computations, simply by conversing
with an AI agent. This forward-looking framework exemplifies how recent
advances in AI can be applied to streamline biological research
workflows, from data processing and visualization to hypothesis testing,
all within a user-friendly, reproducible, and extensible system.

\section{Statement of Need}\label{statement-of-need}

Modern biology and medicine are awash in data, from high-throughput
sequencing to large-scale imaging, and unlocking insights from these
datasets increasingly requires complex computational analyses. However,
a large fraction of life scientists lack the programming expertise to
fully utilize advanced bioinformatics methods. While point-and-click
software and online tools exist for certain tasks, they are limited in
scope and flexibility. In practice, many biologists face a steep
learning curve to perform custom analyses, often having to either master
scripting or rely on expert collaborators. This accessibility gap slows
down research progress and creates silos between experimentalists and
computational experts. There is a clear need for tools that lower the
barrier to bioinformatics, enabling domain experts to directly interact
with their data and ask sophisticated questions without writing code
themselves.

Large language models offer a compelling solution to this problem: they
can understand natural language and produce workable code. In principle,
an LLM-powered assistant could allow a biologist to simply describe the
analysis they want, and receive results with minimal manual effort.
General-purpose AI chatbots (e.g., ChatGPT) have indeed demonstrated the
ability to generate code snippets {[}2{]}. Yet, on their own they fall
short as practical lab tools. For example, a biologist might ask ChatGPT
to ``load my single-cell RNA-seq data and find marker genes for each
cluster,'' and ChatGPT can suggest Python code using libraries like
Scanpy. But it cannot actually execute that code or verify that it
worked on the researcher's data. Moreover, it has no direct access to
local data files or scientific file formats; a user cannot simply feed a
raw .h5ad file into ChatGPT and expect a result, because the model can't
parse that binary format or handle large datasets within its text-only
interface. These limitations mean that, in practice, scientists would
still need to copy the AI-generated code into their own environment,
install the right libraries, troubleshoot errors, and interpret
outputs---negating much of the promised convenience {[}3{]}.

OLAF is designed to fulfill this unmet need by providing an end-to-end
platform where an LLM agent not only writes code but also executes it in
an integrated bioinformatics environment. This design makes advanced
analyses accessible to researchers who are not proficient coders. The
target audience includes bench scientists, clinicians, and students in
life sciences who want to perform data analysis or pipeline tasks by
simply describing their goals. For instance, an immunologist could
upload an .h5ad file containing single-cell transcriptomics data and
ask, ``What cell types are present and what are their top marker
genes?'' OLAF's backend would dispatch this request to its LLM agent,
which might generate code to read the file with Scanpy {[}1{]}, run
clustering and marker gene identification, then plot a UMAP colored by
predicted cell type. The platform would run this code automatically and
return interactive plots and summaries to the user. In this scenario,
OLAF removes the need for the researcher to manually code or even know
the names of the required libraries---the AI agent bridges that gap.
This dramatically lowers the technical hurdle for performing
sophisticated analyses, allowing scientists to focus on the biological
questions rather than the computational mechanics.

Beyond analysis, OLAF contributes to lab automation by enabling natural
language control of computational workflows. Researchers can chain
together complex tasks (data preprocessing, analysis, visualization,
etc.) through simple instructions, effectively scripting full pipelines
via conversation. In traditional lab settings, such automation required
writing elaborate scripts or using workflow managers, both of which
demand programming. OLAF offers a more intuitive alternative,
potentially saving time and reducing human error. It also serves an
educational purpose: by exposing the generated code for each step, OLAF
helps users learn how certain analyses are conducted programmatically,
fostering transparency and reproducibility in computational biology
{[}4{]}. This aligns with the open-science ethos, as OLAF is released
under an open-source license for the community. Researchers can extend
or modify the platform---for example, by adding new analysis modules or
adapting the LLM prompts for their specific domain---thereby continually
expanding OLAF's capabilities. In summary, the need for OLAF is evident
in the daily bottlenecks faced by life scientists: it provides an
accessible, automated, and customizable solution for conducting
bioinformatics analyses, filling a critical gap between rapidly evolving
AI technology and its real-world utility in biological research.

\subsubsection{Table 1: Comparison of AI-Driven Tools for Bioinformatics
Workflows}\label{table-1-comparison-of-ai-driven-tools-for-bioinformatics-workflows}

\rowcolorscustom

\begin{longtable}[]{@{}
  >{\raggedright\arraybackslash}p{(\linewidth - 8\tabcolsep) * \real{0.3113}}
  >{\raggedright\arraybackslash}p{(\linewidth - 8\tabcolsep) * \real{0.2517}}
  >{\raggedright\arraybackslash}p{(\linewidth - 8\tabcolsep) * \real{0.1325}}
  >{\raggedright\arraybackslash}p{(\linewidth - 8\tabcolsep) * \real{0.2119}}
  >{\raggedright\arraybackslash}p{(\linewidth - 8\tabcolsep) * \real{0.0927}}@{}}
\toprule\noalign{}
\begin{minipage}[b]{\linewidth}\raggedright
\textbf{Feature / Capability}
\end{minipage} & \begin{minipage}[b]{\linewidth}\raggedright
\textbf{General LLMs (e.g., ChatGPT)}
\end{minipage} & \begin{minipage}[b]{\linewidth}\raggedright
\textbf{Bio-\\Chatter}
\end{minipage} & \begin{minipage}[b]{\linewidth}\raggedright
\textbf{LangChain-Based Agents}
\end{minipage} & \begin{minipage}[b]{\linewidth}\raggedright
\textbf{OLAF}
\end{minipage} \\
\midrule\noalign{}
\endhead
\bottomrule\noalign{}
\endlastfoot
Understands natural language & Y & Y & Y & \textbf{Y} \\
Generates domain-specific code & L & Y & Y & \textbf{Y} \\
Executes code automatically & N & N & L & \textbf{Y} \\
Handles scientific file formats (e.g., .h5ad) & N & L & L &
\textbf{Y} \\
Uploads and accesses local data files & N & L & Y & \textbf{Y} \\
Provides interactive data visualizations & N & N & L & \textbf{Y} \\
Designed for non-programmers & N & L & N & \textbf{Y} \\
End-to-end analysis workflows & N & L & L & \textbf{Y} \\
Modular and extensible architecture & N & Y & Y & \textbf{Y} \\
Built-in execution environment (sandboxed) & N & N & L & \textbf{Y} \\
Educational transparency (code + outputs shown) & N & Y & Y &
\textbf{Y} \\
Open-source and community-driven & N & Y & Y & \textbf{Y} \\
\end{longtable}

\subsubsection{Key}\label{key}

\begin{itemize}
\tightlist
\item
  \textbf{Y} = Yes (Fully Supported)
\item
  \textbf{L} = Limited Support
\item
  \textbf{N} = No Support
\end{itemize}

\section{State of the Field}\label{state-of-the-field}

The introduction of powerful LLMs like GPT-4 {[}5{]} has sparked a wave
of innovation in computational biology, with researchers exploring how
these models can assist in coding, data analysis, and knowledge
discovery. Early successes in using LLMs for scientific purposes have
mostly revolved around natural language tasks. For example, BioGPT
{[}6{]} and similar models have been used for literature mining and
biomedical question-answering. Code generation for scientific computing
is a newer frontier that has gained traction in the last two years.
Researchers have begun to assess how well general LLMs can write
bioinformatics code: for instance, the BioCoder benchmark {[}7{]}
evaluates models on tasks like generating functions for sequence
analysis and finds that only the most advanced models (e.g., GPT-4)
possess enough domain knowledge to perform well. These studies highlight
that while off-the-shelf LLMs can produce code, specialized domains
require deeper integration with domain-specific data types and
workflows. Even when LLMs know the correct APIs, their code often
requires iterative debugging to run on real datasets. Thus, a key focus
in this field has been how to integrate LLMs with execution environments
and domain-specific knowledge bases to make them truly useful for
scientists.

A few recent systems and frameworks have pointed the way forward. For
example, ChatGPT's Code Interpreter {[}2{]} demonstrated the benefit of
coupling an LLM with a sandboxed Python environment, allowing users to
upload data files and get real-time results. In the life sciences,
researchers at EMBL introduced BioChatter {[}3{]}, an open-source
framework that allows LLMs to call external tools and databases, showing
how to overcome an LLM's knowledge cutoff and lack of interactivity.
Similarly, projects in lab automation have combined LLM agents with
instrument control: for instance, the LABIIUM system {[}8{]} uses an LLM
assistant to generate code for operating lab equipment, lowering the
barrier for experimental automation. Across these developments, a common
theme is emerging: LLM-centric ``agents'' need a supportive
infrastructure that provides data access, domain-specific tools, and a
structured way to orchestrate multi-step workflows {[}9,10{]}. Simply
having a powerful language model is not enough; it must be embedded in a
broader context that handles the realities of data processing, tool
interoperability, and user interaction.

Despite this progress, the computational biology community currently
lacks a turnkey platform that non-programmers can use for on-demand data
analysis powered by LLMs. Existing solutions tend to be either
proof-of-concept prototypes or lower-level libraries requiring developer
expertise, such as LangChain {[}11{]}, which is more of a toolkit than a
finished product. Until now, a bench scientist interested in using an
LLM for data analysis would have to juggle multiple pieces: one tool to
chat with the LLM, another environment to run any suggested code, and no
simple way to connect them. OLAF distinguishes itself by addressing this
gap: it combines the latest LLM agent techniques with a ready-to-use
interface and execution backend specifically tailored for
bioinformatics. Its agent--pipe--router architecture is informed by best
practices for building code-executing LLMs {[}9{]}. In OLAF, an
LLM-based router first interprets the user's request and decides which
specialized module (agent) or sequence of actions (pipeline) is needed.
The code that is generated is then passed along a ``pipe'' for execution
in a sandboxed environment, and the results can be inspected or even fed
back into subsequent LLM prompts if iterative reasoning is required. By
constraining and structuring the agent's possible actions, OLAF also
improves reliability compared to a completely open-ended AI approach, an
important design principle noted in broader evaluations of language
models {[}10{]}.

Looking forward, we anticipate several exciting directions for OLAF and
similar systems. One immediate trajectory is expanding the repertoire of
analyses and file types OLAF can handle, for example, integrating
modules for imaging data or proteomics. Another important direction is
incorporating domain-specific LLMs or fine-tuned models, which could
significantly boost accuracy on niche bioinformatics tasks. Scaling and
collaboration features are also on the horizon: OLAF could be extended
to run computationally intensive analyses on remote servers or cloud
clusters, letting users handle big datasets through the same simple
interface. Additionally, as lab automation evolves, OLAF might integrate
with laboratory information management systems (LIMS) or even robotic
instruments, bridging wet-lab and dry-lab automation {[}12{]}. For
instance, a future version of OLAF might allow a user to say, ``Analyze
this new sequencing run and prepare a report, then queue the next
sample,'' automatically triggering real-world actions.

Crucially, OLAF is open-source and modular, meaning the community can
actively contribute to these expansions. By being open, it invites
collaboration to add new analytical methods and to share best practices
for AI-assisted research, which is essential in a field moving as
rapidly as AI in biology.

\section{Implementation}\label{implementation}

\begin{figure}
\centering
\pandocbounded{\includegraphics[keepaspectratio]{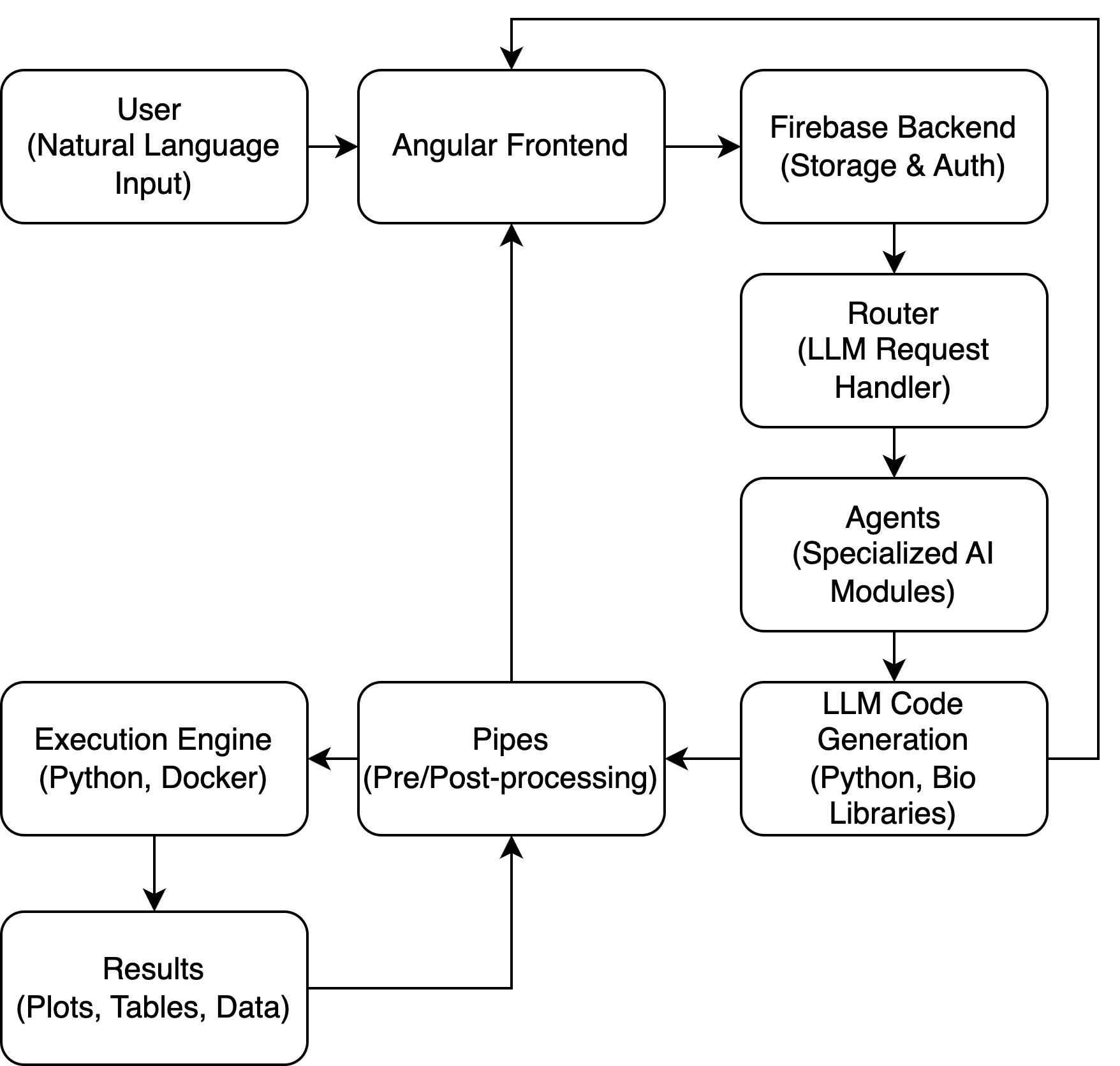}}
\caption{OLAF Architecture}
\end{figure}

\textbf{Figure 1}: The architecture of OLAF, showcasing the interaction
between the front-end (Angular), backend (Python/Firebase), and the
modular agent--pipe--router system. This design ensures scalability,
maintainability, and ease of integration for bioinformatics workflows.

OLAF is split into two major layers. The front-end is built in Angular,
where features such as session management, interactive data uploads, and
real-time result displays reside. Angular's component-based design is
used extensively to ensure a logical separation of views and
data-handling services. The second layer is a Python-driven backend
deployed as Firebase Cloud Functions, orchestrated through Docker for
local testing and reproducibility {[}13{]}.

\textbf{Front-End (Angular)}

The front-end is built using Angular, a robust TypeScript-based web
application framework. OLAF adopts standard Angular best practices,
including:

\begin{itemize}
\tightlist
\item
  Service-based architecture for business logic and API communication
\item
  Strongly typed models to ensure consistency in data exchange
\item
  Component-based views to enable modular design and reuse
\item
  Page-based routing to allow intuitive session navigation and
  multi-step workflows
\end{itemize}

Features such as session management, interactive file uploads, real-time
data visualization, and code/result inspection panels are all
encapsulated as Angular components. Services abstract the communication
between the UI and backend API routes, while models represent key data
types like session metadata, pipeline outputs, and user inputs. This
structure provides not only a maintainable and scalable UI but also a
solid base for future extensions such as collaborative sessions, user
authentication, or domain-specific toolkits.

\textbf{Backend (Python + Firebase + Docker)}

\textbf{Agents}. Each agent encapsulates a distinct AI or transformation
capability. An agent might interact with external APIs (such as OpenAI
or a specialized bioinformatics service) or perform local computations.
By confining decision-making and data transformation logic to dedicated
agent classes, OLAF ensures that specialized tasks are simple to add and
maintain.

\textbf{Routers}. The router layer receives session data from front-end
requests, determines which agent should handle the request, and then
organizes the flow of data among various pipes. If an agent's response
calls for chaining multiple operations, the router re-invokes itself
with the new route. This design makes it clear and explicit how incoming
data is processed, reduces duplication, and simplifies testing by
isolating routing logic from agent logic.

\textbf{Pipes}. Pipes are asynchronous transformations that can be
chained and reused to preprocess or postprocess data. For example, a
pipe might scrub sensitive user input, attach domain-specific metadata,
or handle rate-limiting for external APIs. By separating these
responsibilities into well-defined components, OLAF maintains clarity
and minimizes side effects throughout the processing pipeline.

The overall architecture is illustrated in Figure 2. Because OLAF is
designed to run under Docker locally, users and reviewers can easily
spin up both the front-end (Angular's development server) and the
backend (Firebase Emulator or Docker container) to conduct end-to-end
testing.

\begin{figure}
\centering
\pandocbounded{\includegraphics[keepaspectratio]{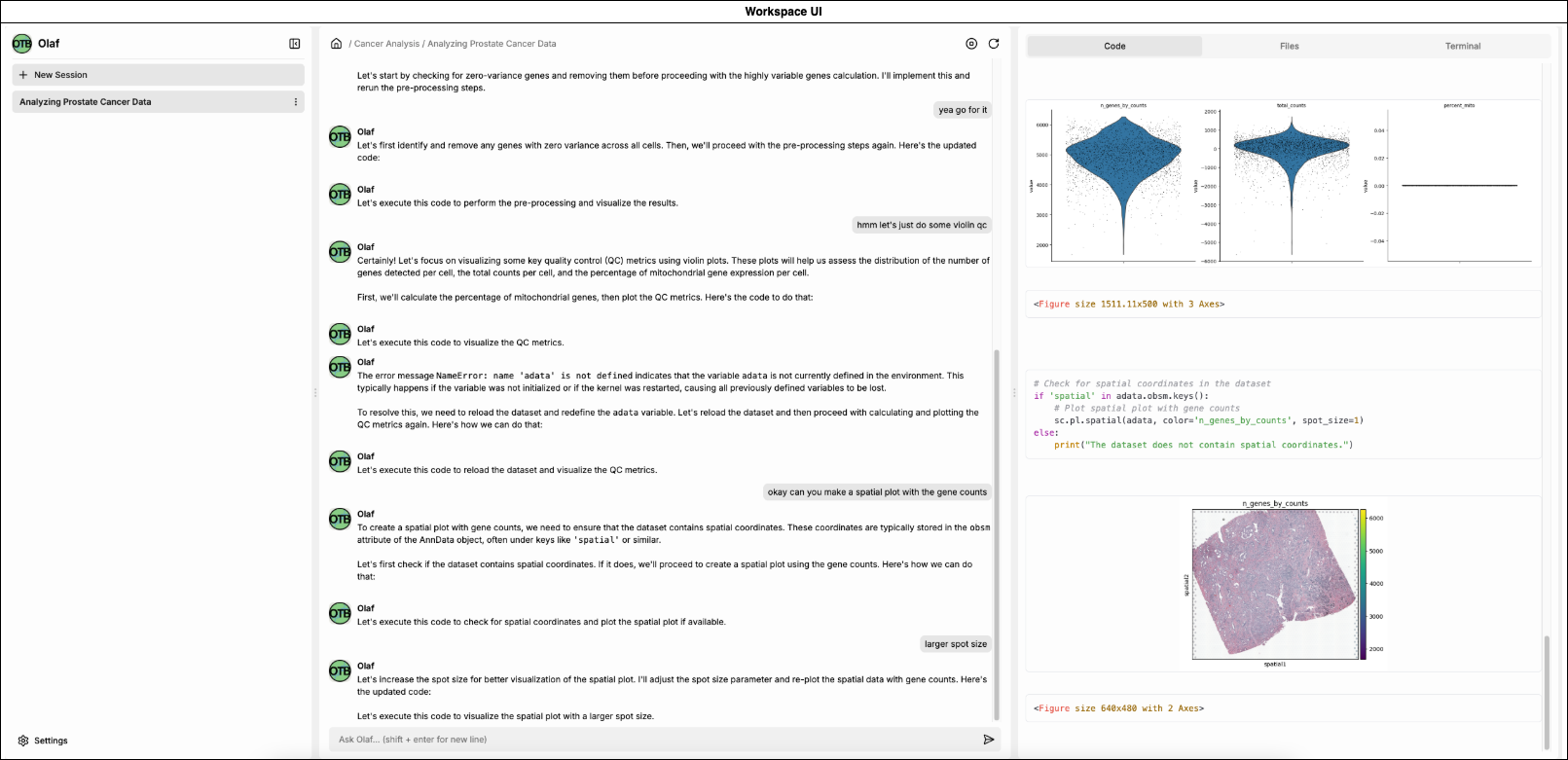}}
\caption{OLAF Webpage Design}
\end{figure}

\textbf{Figure 2}: The OLAF webpage design, illustrating the user
interface layout and interaction flow. This includes features such as
session management, file upload, and real-time visualization, providing
an intuitive and user-friendly experience for bioinformatics workflows.

\section{Code Availability}\label{code-availability}

OLAF is open-source and available at:
\url{https://github.com/OpenTechBio/Olaf}.

The repository includes installation instructions, usage examples, and
documentation to help users get started quickly. Contributions and
issues are welcome via GitHub.

\section{Example Use Case}\label{example-use-case}

In a typical use case, a researcher might engage in a conversation
asking OLAF to perform a series of steps for gene sequence analysis. The
user first provides a short sequence to the system, requesting an
annotation. OLAF's Angular front-end presents this conversation as a
series of chat-like messages. Behind the scenes, the Router receives
session data and dispatches it to a specialized agent capable of
sequence analysis, possibly calling external LLM-based libraries for
advanced annotations (e.g., functional domain predictions or similarity
searches). After the agent computes its responses, data pipes may trim
extraneous symbols or filter untrusted content. If additional steps are
necessary, the MasterAgent can prompt the Router to invoke another route
(for example, a validation step or database write) before returning the
final annotated result to the user's browser. All of this is
accomplished through a simple chat-like interface, lowering the barrier
for domain scientists to run complex computations interactively.

\begin{figure}
\centering
\pandocbounded{\includegraphics[keepaspectratio]{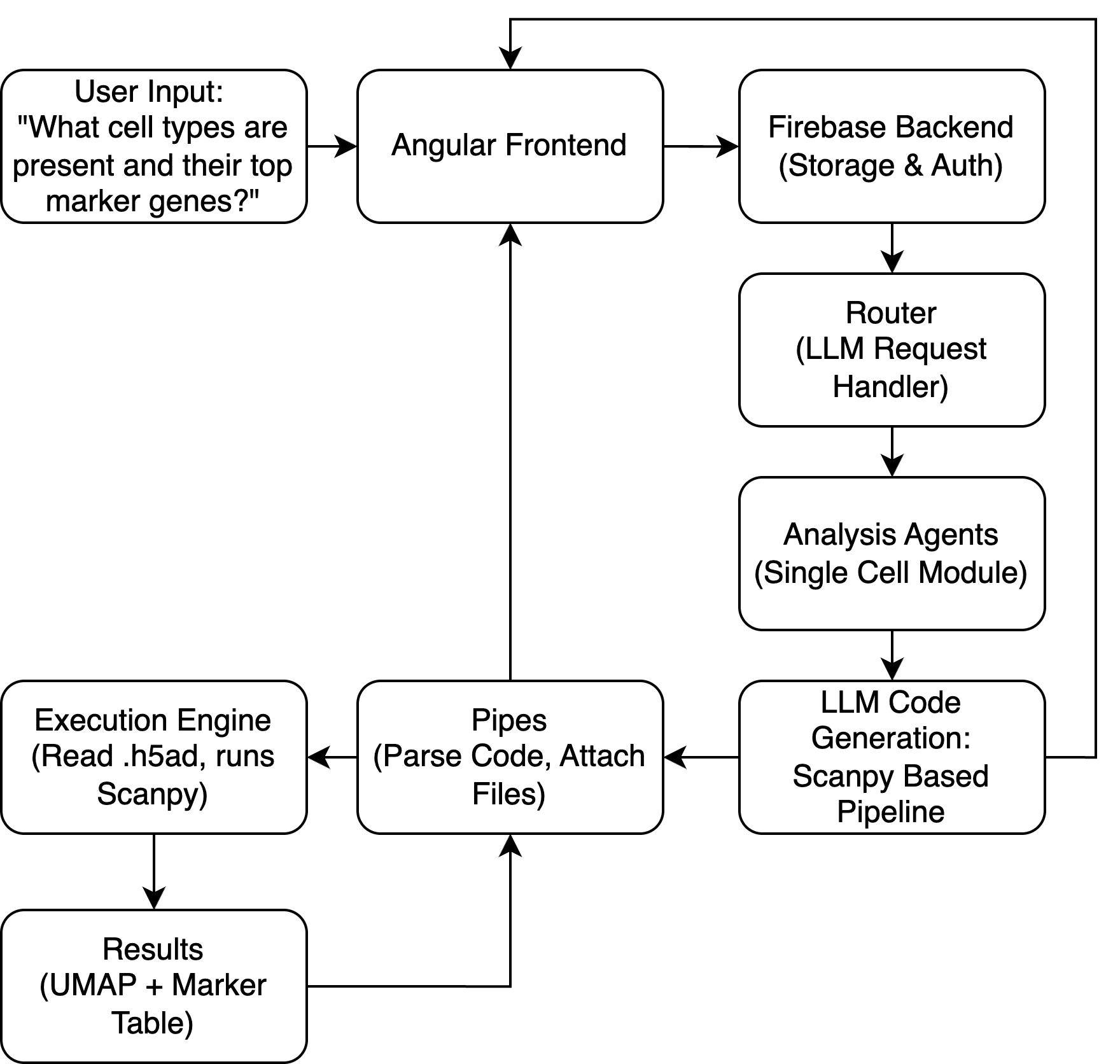}}
\caption{OLAF Use Case}
\end{figure}

\textbf{Figure 3}: Example use case of OLAF in action. The figure
illustrates a typical workflow where a researcher interacts with OLAF to
perform gene sequence analysis. The chat-like interface simplifies the
process by allowing natural language queries, while the backend handles
complex computations and returns annotated results seamlessly.

\section{Future Directions}\label{future-directions}

Future work will present rigorous evaluations of OLAF's performance
across common bioinformatics tasks, including benchmarking code
correctness, user experience, and comparison to existing workflows.
These results, along with biological case studies, are the subject of a
forthcoming manuscript.

\section{References}\label{references}

{[}1{]}: Wolf, F. A., et al.~(2018). Scanpy: Large-scale single-cell
gene expression analysis. Genome Biology.
https://doi.org/10.1186/s13059-017-1382-0

{[}2{]}: OpenAI. (2023). ChatGPT Code Interpreter.
https://openai.com/blog/chatgpt-plugins

{[}3{]}: Lobentanzer, S., et al.~(2022). A platform for the biomedical
application of large language models. Nature Biotechnology.
https://doi.org/10.1038/s41587-024-02534-3

{[}4{]}: Wilkinson, M. D., et al.~(2016). The FAIR Guiding Principles.
Scientific Data. https://doi.org/10.1038/sdata.2016.18

{[}5{]}: OpenAI. (2023). GPT-4 Technical Report.
https://openai.com/research/gpt-4

{[}6{]}: Luo, R., et al.~(2021). BioGPT: Generative Pre-trained
Transformer for Biomedical Text Generation and Mining. Briefings in
Bioinformatics. https://doi.org/10.1093/bib/bbac409

{[}7{]}: Belrose, J., et al.~(2023). BioCoder: A Benchmark for
Bioinformatics Code Generation with Large Language Models.
Bioinformatics. https://doi.org/10.1093/bioinformatics/btae230

{[}8{]}: Olowe, E., et al.~(2024). LABIIUM: AI-Enhanced
Zero-configuration Measurement Automation System. arXiv.
https://doi.org/10.48550/arXiv.2412.16172

{[}9{]}: Lyu, C., et al.~(2024). Large Language Models as Code
Executors: An Exploratory Study. arXiv.
https://doi.org/10.48550/arXiv.2410.06667

{[}10{]}: Liang, P., et al.~(2022). Holistic Evaluation of Language
Models. arXiv. https://doi.org/10.48550/arXiv.2211.09110

{[}11{]}: LangChain Documentation. (2025). LangChain Framework.
https://www.langchain.com

{[}12{]}: Al Kuwaiti, A., et al.~(2023). A Review of the Role of
Artificial Intelligence in Healthcare. Journal of Personalized Medicine.
https://doi.org/10.3390/jpm13060951

{[}13{]}: Merkel, D. (2014). Docker: Lightweight Linux Containers for
Consistent Development and Deployment. Linux J.

\end{document}